\documentclass[11pt]{article}
\usepackage[latin1]{inputenc}
\usepackage{graphicx}
\usepackage{epsfig}
\usepackage{color}
\usepackage{fullpage}
\usepackage{setspace}
\usepackage{lscape}
\usepackage{verbatim}
\usepackage{amsthm, amssymb}
\usepackage{amsmath}
\usepackage{hyperref}

\usepackage[noend]{algpseudocode}
\usepackage{algorithm}
\usepackage{algorithmicx}

\usepackage{xspace}
\usepackage{comment}
\usepackage[backgroundcolor=white]{todonotes}
\newtheorem{Lem}{Lemma}
\newtheorem{Thm}{Theorem}
\newtheorem{Cor}{Corollary}
\newtheorem{Def}{Definition}

\def\polylog{\operatorname{polylog}}

\newcommand{\dist}[2]{\ensuremath{\normalfont{\textbf{d}}_{#1}({#2})}}
\newcommand{\distT}[2]{\ensuremath{\tilde{\normalfont{\textbf{d}}}_{#1}({#2})}}
\newcommand{\VV}{\ensuremath{V^\circ}}
\newcommand{\patterntree}{\ensuremath{\mathcal{A}}}

\makeatletter
\def\algbackskip{\hskip-\ALG@thistlm}
\makeatother

\title{Truly Subquadratic Exact Distance Oracles with Constant Query Time for Planar Graphs}
        \author{Viktor Fredslund-Hansen
        \footnote{Department of Computer Science,
                  University of Copenhagen,
                  \texttt{viha@di.ku.dk}}
        \and Shay Mozes
        \footnote{Efi Arazi school of Computer Science,
        IDC Herzliya, Israel.
        \texttt{smozes@idc.ac.il}
        \texttt{https://cs.idc.ac.il/$_{\widetilde{~}}$smozes/}. Partially supported by Israel Science Foundation grant no. 592/17.}
        \and Christian Wulff-Nilsen
        \footnote{Department of Computer Science,
                  University of Copenhagen,
                  \texttt{koolooz@di.ku.dk},
                  \texttt{http://www.diku.dk/$_{\widetilde{~}}$koolooz/}. This research is supported by the Starting Grant 7027-00050B from the Independent Research Fund Denmark under the Sapere Aude research career programme.}}

\date{}
\begin{document}

\maketitle
\begin{abstract}
Given an undirected, unweighted planar graph $G$ with $n$ vertices, we present a truly subquadratic size distance oracle for reporting exact shortest-path distances between any pair of vertices of $G$ in constant time. 
For any $\varepsilon > 0$, our distance oracle takes up $O(n^{5/3+\varepsilon})$ space and is capable of answering shortest-path distance queries exactly for any pair of vertices of $G$ in worst-case time $O(\log (1/\varepsilon))$. Previously no truly sub-quadratic size distance oracles with constant query time for answering exact all-pairs shortest paths distance queries existed.
\end{abstract}
\thispagestyle{empty}

\newpage

\setcounter{page}{1}

\ifdefined\fullver
\section{Introduction}\label{sec:Intro}
Efficiently answering shortest path distance queries between pairs of vertices in a graph is a fundamental algorithmic problem with numerous important applications. Given an $n$-vertex graph $G=(V,E)$ a \textit{distance oracle} is a data-structure capable of efficiently answering shortest path distance queries between pairs of vertices $u,v \in V$. By ``efficiently'', we usually mean answering distance queries in constant or poly-logarithmic time in the input size while keeping the space usage as low as possible. A naive approach to solving the problem would be to represent the distances by an $n \times n$ distance matrix, as this allows for answering distance queries in constant time by looking up the entry corresponding to the pair. The obvious drawback of this approach, however, is the large space requirement of $\Theta(n^2)$ which may be impractical for larger graphs. The other extreme is to lazily compute a shortest path distance from scratch with each query, for instance with Dijkstra's algorithm. This approach only needs memory for storing the input graph and has no requirement for preprocessing. However, the high query time makes this approach infeasible for many applications where a small query time is of importance.

\subsection{Known lower bounds}
It is well known that there are graphs for which no distance oracle with $o(n^2)$ bits of space and $O(1)$ query time exists. More generally, Thorup and Zwick showed that for any $k\in\mathbb N$, there are dense graphs for which any distance oracle with \textit{stretch} at most $2k-1$ and constant query time requires $\omega(n^{1+1/k})$ bits of space\cite{Thorup2005}; this is conditioned on the widely believed \textit{girth conjecture} of Erd\H{o}s \cite{Erdos1969} which has been proven for small values of $k$. In the context of distance oracles, \textit{stretch} refers to the multiplicative approximation error of the reported distances, that is, a distance oracle which reports an estimate $\distT{G}{u,v}$ for which $\distT{G}{u,v} \leq t \cdot \dist{G}{u,v}$ is said to have stretch $t$. We refer to a distance oracle with stretch $t = 1$ as an \textit{exact distance oracle} and as an \textit{approximate distance oracle} otherwise. Furthermore P\v{a}tra\c{s}cu and Roditty \cite{Patrascu2014} showed that there are sparse graphs on $O(n \polylog n)$ edges for which constant query-time distance oracles with stretch less than $2$ must be of size $\Omega(n^2 \polylog n)$, assuming the set intersection conjecture.
\todo[inline]{VH: Max said that I am wrong about the above lower bound by Patrascu and Roditty - I checked but can not see where I am wrong, but maybe you can verify if I overlooked something?}

\subsection{Distance oracles for planar graphs}
The lower bounds mentioned above provides evidence that the trivial lookup table data structure is essentially the best possible for constant query time in general graphs. It is therefore natural to consider the study of distance oracles in more restricted settings, for instance for certain classes of graphs or when we relax the requirement that exact distances be reported. This paper concerns itself with the special case of exact distance oracles for planar graphs. To the knowledge of the authors there are no non-trivial lower bounds for distance oracles for planar graphs,\todo{Shay: with the exception of a conditional lower bound for dynamic oracles} and thus the ``holy grail'' for distance oracles for planar graphs would be a distance oracle of linear size with constant query time. In the following we provide notable as well as recent developments for distance oracles for planar graphs. 
The setting of planar graphs is also well-motivated as they arise in many important real-world applications, notably in routing, navigation of road and sea maps as well as in the context of computational geometry. \todo{CW: citations mentioning such applications?}

\subsection{Subquadratic-space distance oracles for planar graphs}
Many constructions in the literature of distance oracles for planar graphs provide tradeoffs between query time and size~\cite{Djidjev1996, Arikati1996, Cabello2012, Chen2000, Fakcharoenphol2006, Mozes2012, Nussbaum2011}. However, all of these distance oracles use essentially quadratic space for polylogarithmic query time. As a matter of fact, it was, until recently, a major open problem whether a size $O(n^{2-\varepsilon})$ exact distance oracle with poly-logarithmic query-time could be constructed for an $n$-vertex planar graph where $\varepsilon > 0$ is a constant. This was answered in the affirmative by Cohen-Addad et al.~\cite{Cohen-Addad2017} who presented a construction of size $O(n^{5/3})$ distance oracle with query time $O(\log n)$, as well as a space/query-time tradeoff allowing for construction of size $S$ distance oracles with query time $O\left((n^{5/2}(S^{3/2}) \log n \right)$ for $S \geq n^{3/2}$. Their results were inspired by the novel \textit{abstract Voronoi diagram} techniques introduced by Cabello \cite{Cabello2018} in his breakthrough result, where he gave the first truly sub-quadratic time algorithm for computing the diameter of planar graphs. By devising an elegant point-location structure for the abstract Voronoi diagrams, Gawrychowski et al.~\cite{Gawrychowski2018} managed to further improve on this bound: They presented a size $O(n^{3/2})$ distance oracle with $O(\log n)$ query time, as well as a tradeoff allowing for size $O(S)$ distance oracles with query time $O(\max \left\{ 1, n^{3/2} \polylog n/S \right\})$ for $n \leq S \leq n^2$. Very recently, Charalampopoulos et al.~\cite{Charalampopoulos2019a} presented constructions that are much closer to the holy-grail. 
They described the following oracles for edge-weighted planar digraphs; one of size $O(n^{1+\varepsilon} \polylog n)$ with $O(\log^{1/\varepsilon} n)$ query time, another of size $O(n \polylog n)$ with $O(n^{\varepsilon}  \polylog n)$ query time and one of size $n^{1+o(1)}$ with query-time $n^{o(1)}$ for constant $\varepsilon > 0$. They achieved this by combining the very same point location structure of \cite{Gawrychowski2018} with a clever recursion scheme. It should be emphasized that all of the truly sub-quadratic-size exact distance oracles so far require at least poly-logarithmic query time. 
Considering oracles with constant query time, the biggest improvement in space over the trivial look-up table data structure result is less than a $\log n$-factor even when restricted to unweighted undirected planar graphs~\cite{Wulff-Nilsen2013, Chan12, Wulff-Nilsen2010}.

\subsection{Approximate distance oracles for planar graphs}
In spite of the absence of lower bounds, the lack of progress in sparse distance oracles when the query time must be constant (or poly-logarithmic until recently) has motivated the study of approximate distance oracles for restricted classes of graphs. Klein presented a size $O(\varepsilon^{-1} n \log n)$ distance oracle for undirected planar graphs capable of reporting $(1+\varepsilon)$-approximate distances in time $O(\varepsilon^{-1})$ \cite{Klein2002}. This result was essentially a simplified and streamlined version of Thorup's size $O(\varepsilon^{-1} n (\log n)(\log(n)))$ compact oracle for planar digraphs capable of reporting $(1+\varepsilon)$-approximate distances in time $O(\log \log(n) + \varepsilon^{-1})$ per query\footnote{This construction also generalized to a size $O(\varepsilon^{-1} n (\log n))$ construction undirected graphs.} \cite{Thorup2004}.

The first truly linear-size construction was due to Kawarabayashi et al. who presented a size $O(n)$ distance oracle for undirected planar graphs \cite{Kawarabayashi2011}. The oracle is capable of reporting $(1+\varepsilon)$-approximate distances in (slightly larger) time $O(\varepsilon^{-2} \log^2 n)$. We note that their constructions also generalize to bounded-genus, and minor-free graphs. A subsequent result by Kawarabayashi et al. aimed to reduce the dependency on $\varepsilon^{-1}$, presenting the first construction in which space-query product is $\tilde{o}(n \varepsilon^{-2})$, giving a size $\tilde{O}(n)$ distance oracle capable of reporting $(1+\varepsilon)$-approximate queries in time $\tilde{O}(\varepsilon^{-1})$. For fixed $\varepsilon$, Wulff-Nilsen improved the space-query of the constructions of Thorup, respectively Klein, from $O(n \log n)$ to $O(n (\log \log n)^5)$ \cite{Wulff-Nilsen2016}.

Gu and Xu presented a size $\tilde{O}(n)$ distance oracle capable of reporting $(1+\varepsilon)$-approximate distances in time $O(1)$. This is the first oracle with query time independent of $\varepsilon$ and with near-linear preprocessing time, subsuming the constructions whose query times are linear in $\varepsilon^{-1}$. This should also be compared with the exact oracle of Wulff-Nilsen for which their construction is of smaller size for various choices of $\varepsilon$ \cite{Wulff-Nilsen2010}. The preprocessing time and space of this construction was recently improved by Chan and Skrepetos who replaced an exponential dependency on $\varepsilon^{-1}$ with a polynomial dependency and furthermore shaving off a factor $\log n$ from the preprocessing \cite{Chan2019}, while increasing the query time to $O(\log{\varepsilon^{-1}})$.

For a less recent, but more comprehensive treatment of developments and techniques employed in the field of distance oracles, we refer to the survey of Sommer \cite{Sommer2014}.

\else
\section{Introduction}\label{sec:Intro}
Efficiently answering shortest path distance queries between pairs of vertices in a graph is a fundamental algorithmic problem with numerous important applications. Given an $n$-vertex graph $G=(V,E)$ a \textit{distance oracle} is a compact data-structure capable of efficiently answering shortest path distance queries between pairs of vertices $u,v \in V$.
Ideally one would like the data structure to be of linear size and the query time to be constant. However, it is well known that there are graphs for which no distance oracle with $o(n^2)$ bits of space and $O(1)$ query time exists. In fact, even resorting to approximation, P\v{a}tra\c{s}cu and Roditty \cite{Patrascu2014} showed that there are sparse graphs on $O(n \polylog n)$ edges for which constant query-time distance oracles with stretch less than $2$ must be of size $\Omega(n^2 \polylog n)$, assuming the set intersection conjecture. These impossibility results make it natural to consider the probelms in restricted classes of graphs.

In this paper we consider exact distance oracles for planar graphs. Distance oracles for planar graphs are well motivated by important real-world applications,  
notably in routing, navigation of road and sea maps as well as in the context of computational geometry. 
To the best of our knowledge there are no non-trivial lower bounds for (static) distance oracles for planar graphs, 
and thus achieving the ``holy grail'' of a linear-size distance oracle with constant query time may be possible.
Indeed, there has been numerous works over at least three decades developing exact distance oracles for planar graph~\cite{Djidjev1996, Arikati1996, Cabello2012, Chen2000, Fakcharoenphol2006, Mozes2012, Nussbaum2011}. However, only recently Cohen-Addad et al.~\cite{Cohen-Addad2017} gave the first oracle with truly subquadratic space and polylogarithmic query time. 
Their result was inspired by Cabello's \cite{Cabello2018} breakthrough result, who gave the first truly sub-quadratic time algorithm for computing the diameter of planar graphs by a novel use of Voronoi diagrams. 
The approach of~\cite{Cohen-Addad2017} was subsequently improved by~\cite{Gawrychowski2018,Charalampopoulos2019a}, who gave an elegant point-location mechanism for Voronoi diagrams in planar graphs, and combined it with a clever recursive scheme to obtain exact distance oracles for directed weighted planar graphs with $O(n^{1+\varepsilon} \polylog n)$ space and $O(\log^{1/\varepsilon} n)$ query time for any small constant $\varepsilon$. 
We note that even though the oracle of~\cite{Charalampopoulos2019a} gets quite close to optimal, it remains wide open to support exact queries in \emph{constant time} using truly subquadratic space, even in the most basic case of unweighted undirected planar graphs~\cite{Wulff-Nilsen2013, Chan12, Wulff-Nilsen2010}.

Allowing approximate answers does help in planar graphs. Many results reporting $(1+\varepsilon)$-approximate distances with various tradeoffs exist, all with (nearly) linear size and polylogarithmic, or even $O(1/\varepsilon)$ query-time~\cite{Thorup2004,Klein2002,Kawarabayashi2011,Wulff-Nilsen2016}. 
Gu and Xu\cite{GuX15} presented a size $O(n\polylog n)$ distance oracle capable of reporting $(1+\varepsilon)$-approximate distances in time $O(1)$. While their query time is a constant independent of $\varepsilon$, the preprocessing time and space are nearly linear, but  with an exponential dependency on $(1/\varepsilon)$. This exponential dependency was recently improved to polynomial~\cite{Chan2019}.

Thus, despite the large body of work on distance oracles for planar graphs, it has remained an open question to determine whether an exact distance oracle with of size $O(n^{2-\varepsilon})$ with \textit{constant} query-time can be constructed for some constant $\varepsilon > 0$.
\fi

\paragraph{Our results and techniques.}
\ifdefined\fullver
It has thus remained an open question whether an exact distance oracle with of size $O(n^{2-\varepsilon})$ with \textit{constant} query-time can be constructed for some constant $\varepsilon > 0$.
\fi
 We answer this question in the affirmative. Our result is presented in the following theorem:
\begin{Thm}\label{thm:main}
Let $G=(V,E)$ be an undirected unweighted $n$-vertex planar graph. For any $\varepsilon > 0$ there exists a data-structure requiring $O(n^{5/3+\varepsilon})$ space that, for any $s,t \in V$, reports the shortest path distance between $s$ and $t$ in $G$ in time $O(\log(1/\varepsilon))$.
\end{Thm}

We remark that the simple distance oracle we present in Section~\ref{sec:construction-1} can be distributed into a distance labeling scheme with size $O(n^{3/4})$ per label, such that the distance between any two vertices $s,t$ can be computed in $O(1)$ time given just the labels of $s$ and $t$.

The main concept we use to obtain our result is that of a \emph{pattern} capturing distances between a vertex and a cycle. This concept was used by \cite{Wulff-Nilsen2013} , and weas called ``distance tuple'' by
and \cite{Li2019}. Consider a vector storing the distances from a vertex $u$ to the vertices of a cycle $\beta$ in their cyclic order. 
The pattern of $u$ w.r.t. $\beta$ is simply the discrete derivative of this vector. That is, the vector obtained by taking the difference between every pair of consecutive values. Li and Parter~\cite{Li2019} observed that when the input graph is planar, the number of different patterns w.r.t. a face with $r$ vertices is $O(r^3)$ regardless of the size of the graph. We next outline how this observation can be used to break the quadratic space barrier.

Roughly speaking, any planar graph can be decomposed into $O(n/r)$ subgraphs, called regions, of size $r$ each, where the boundary of each region (i.e., the vertices that have neighbors outside the region) is a single cycle $h$.\footnote{In fact, a constant humber of cycles. To readers familiar with the concept, this is just an $r$-division with a few holes, but \emph{without} the important feature that each region has just $O(\sqrt r)$ boundary vertices. This is because one cannot triangulate unweighted graphs without changing the distances.}  
Applying Li and Parter's observation in this setting, the number of different patterns for the hole of each region $R$ is $O(r^3)$. Hence, we can represent the distances from any vertex $s \notin R$ to $h$ by just storing the distance from $s$ to an arbitrarily chosen canonical vertex $v_h$ of $h$, and a pointer to the pattern of $s$ with respect to $h$. This requires just $O(n)$ space plus $O(r^3 \cdot r)$ for storing all the patterns for $h$. Summing over all $O(n/r)$ regions, the space required is $O(n^2/r + nr^3)$.
We define the notion of distance from a pattern to a vertex (see Definition~\ref{def:p2v}). While this definition is simple, it is somewhat unnatural because the distance from a pattern to a vertex does not necessarily correspond to the length of any specific path in the graph! However, the distance between $s$ and any vertex $t\in R$ is just the sum of the distance between $s$ and the canonical vertex $v_h$ and the distance from the pattern of $s$ with respect to $h$ to $t$.

We therefore store the distances from each of the $O(r^3)$ possible patterns of $R$ to each vertex of $R$.
This requires $O(r^3 \cdot r)$ space per region, so $O(nr^3)$ space overall. This way we can report the distance between $s$ and $t$ in constant time by (i) retrieving the pattern $p$ of $s$ w.r.t. $h$, and (ii) adding the distance from $s$ to the canonical vertex $v_h$ of $h$ and the distance from the pattern $p$ to $t$. 
These ideas alone already imply an oracle with space $\tilde O(n^{7/4})$ and constant query time. Combining these ideas with recursion yields the improved space bound of Theorem~\ref{thm:main}.

As we argued in the introduction, breaking the quadratic space barrier for constant query time is important and significant result in its own right. We highlight the following difference between the approach taken in our recursive oracle and the approaches used in all existing distance oracles we are aware of. 
To the best of our knowledge, all existing distance oracles, both exact and approximate, and both for general graphs and planar graphs, recover the distance from $s$ to $t$ by identifying a vertex or vertices on some (possibly approximate) shortest path between $s$ and $t$, for which distances have been stored in the preprocessing stage. These vertices are usually referred to as landmarks, portals, hubs, beacons, seeds, or transit nodes~\cite{Sommer2014}.  
Our oracle, on the other hand, reports the exact shortest path without identifying vertices on the shortest path from $s$ to $t$. Instead, it "zooms in" on $t$ by recovering distances to the canonical vertices of a sequence of subgraphs of decreasing size that terminates at a constant size graph containing $t$. We emphasize that \emph{none of these canonical vertices necessarily lies on a shortest path} from $s$ to $t$. This property may be viewed as a disadvantage if we also want to report the shortest path, but when reporting multiple edges on long paths, constant query time is no longer relevant. On the other hand, it may be that just reporting the distance is easier than also reporting an internal vertex on a shortest path. Hence, it may be that developing oracles based on this new approach may lead to further advances on the way to linear size distance oracles for planar graphs with constant query time, and in other related problems.

\ifdefined\fullver
\subsection{Organization}
Our paper is organized as follows: Preliminaries are given in Section~\ref{sec:defs-notation}. Section~\ref{sec:patterns} defines patterns and their properties.
Our main result is presented by going via two intermediate constructions in Sections~\ref{sec:construction-1} and~\ref{sec:construction-2}, respectively. These constructions serve as a warm-up and incorporate techniques from the distance oracle of \cite{Wulff-Nilsen2013} of compressing distances using $r$-divisions and the metric compression framework of Li and Parter \cite{Li2019} \todo{VH: Two constructions altogether?}. Our second construction is similar to the first, but improves the space usage by using recursive planar decompositions instead of the flat $r$-divsion combined with a recursive distance querying scheme somewhat akin to that of \cite{Charalampopoulos2019a}. Our main result is then obtained in Section~\ref{sec:construction-3} by a further refinement that allows fewer distances to be stored explicitly. An important new tool here is a prefix-sum data structure that allows us to compactly represent sums of differences of certain distances; this tool is presented in Section~\ref{sec:prefix-sum-data}.
To simplify our presentation, some technical details have been omitted in the Sections mentioned above. These details can be found in Section~\ref{sec:technicalities}.
\fi

\section{Preliminaries}\label{sec:defs-notation} 
Let $G$ be a graph. We denote by $V(G)$ and $E(G)$ the vertex and edge-set of $G$, and denote by $n = |V(G)|$ the number of vertices of $G$. 
\ifdefined\fullver
A connected component of an undirected graph $H$ is a maximum-size vertex subset $C \subseteq V(H)$ s.t.~$H[C]$ is connected.
A \textit{subgraph} $H \subseteq G$ is a graph $H$ for which $E(H) \subseteq E(G)$ and $V(H) \subseteq V(G)$. The \textit{vertex-induced subgraph} of $G$ is a subgraph of $G$ consisting of all vertices of some vertex-subset and all edges connecting pairs of vertices of that subset in $G$. Similarly, an \textit{edge-induced subgraph} of $G$ is a subgraph of $G$ consisting of some edge-subset and all their vertices to which they are adjacent in $G$. 
\fi
For a subset $S$ of edges or vertices we denote by $G[S]$ the  subgraph of $G$ induced on $S$.
We denote by $u \leadsto_H v$ a \textit{shortest path} from $u$ to $v$ in the subgraph $H$, by $\dist{H}{u,v}$ the length of $u \leadsto_H v$, and define $u \leadsto v \equiv u \leadsto_G v$. 

The following definitions will be useful when talking about decompositions of $G$. A \textit{region} $R$ of $G$ is an edge-induced subgraph of $G$, and its \textit{boundary} $\partial R$ is the vertices of $R$ that are adjacent to some vertex of $V(G) \setminus V(R)$ in $G$. Vertices of $V(R)\setminus\partial R$ are called \textit{interior} vertices of $R$. 
Observe that for a region $R$ and for $u \in R$ and $v \in V \setminus V(R)$, any 
path from $u$ to $v$ in $G$ must intersect $\partial R$.

\ifdefined\fullver 
\paragraph{Planar graphs and embeddings:} 

An \textit{embedding} of a graph $G$ assigns to each vertex a point in the plane and to each edge a simple curve s.t.~its endpoints coincide with those of the points assigned to its vertices. A \textit{planar embedding} of $G$ is an embedding such that no two vertices are assigned the same point and such that no two curves coincide in points other than those corresponding to vertices they share. A graph is said to be \textit{planar} iff it admits a planar embedding. When we talk about a planar graph we assume that it is \textit{planar embedded} and hence that there is some implicit, underlying planar embedding of the graph. When it is clear from context, we shall refer interchangeably to a planar graph and its embedding, to its edges and curves and to its vertices and points.
\fi 

It will be useful to assume some global strict order on a vertex set $V$ s.t.~for any $U \subseteq V$ there is a minimum vertex $\min U \in U$ w.r.t this order. We refer to this as the \emph{canonical vertex} of $U$.

\paragraph{Faces and holes:} 
\ifdefined\fullver
\else
We assume the reader is familiar with the the basic definitions of planarity and of planar embeddings.
\fi
The edges of a plane graph induce maximal open portion of the plane that do not intersect any edges. 
A \textit{face} of the graph is the closure of one such portion of the plane. 
We refer to the edges bounding a face as the \textit{boundary} of that face. Given a face $f$, $V(f)$ is the set of vertices on the boundary of $f$. We denote by $w(f)$ the \textit{facial walk} of $f$ which is the sequence of vertices encountered when walking along $f$ starting at $\min V(f)$ and going in the clockwise direction. Note that $f$ may be non-simple, so some vertices may appear multiple times in $w(f)$. 

A \textit{hole} $h$ in a region $R$ of a graph $G$ is a face of $R$ which is not a face in $G$. 
We say that a vertex $u \in V(G) \setminus V(R)$ is \emph{inside} hole $h$ if $u$ lies in the region of the plane corresponding to the face $h$ of $R$. 
We denote by 
$\VV(h) = \left\{ u\ \in V(G) \; | \; u \text{ is inside } h \right\}$ 
all the vertices that are inside $h$.

\ifdefined\fullver
\subsection{Separators, planar decompositions and $r$-divisions}\label{sec:separators}
A \textit{vertex separator} of an undirected graph $H$ is a subset of vertices of $V(H)$ that decomposes $H$ into two subgraphs s.t.~any path with endpoints in different subgraphs intersects the vertex separator. A \textit{balanced vertex separator} is a vertex separator where each of its subgraphs contain at most a constant fraction of the vertices of the graph it separates. Often it is also desirable that the size of the separator is small. A \textit{cycle separator} is a vertex separator whose vertex-induced subgraph~\todo{CW: does this assume that the graph is triangulated?} is a simple cycle.

\paragraph{$r$-divisions of planar graphs}
For any $n$-vertex planar graph $G$, Frederickson \cite{Frederickson1987} showed that by carefully applying the planar vertex-separator theorem of Lipton and Tarjan \cite{Lipton1979} one can obtain a so-called \textit{$r$-division} which is a decomposition of $G$ into $O(n/r)$ regions s.t. each region has at most $O(r)$ vertices and $O(d\sqrt{r})$ boundary vertices, where $d$ is the size of the largest face. Our constructions rely heavily on such decompositions, and for some applications, including ours, it is required that the regions of the $r$-division furthermore behave well from a topological point-of-view, i.e. they have a ``few'' (i.e. constant) number of holes (since recursively partitioning the input graph until each region has size $O(r)$ causes holes to appear in regions).  Such an $r$-division with a constant number of holes can be obtained by a repeated application of the cycle separator theorem of Miller \cite{Miller1986}, but instead of always balancing the number of vertices in the resulting subgraphs, the number of holes and boundary nodes in a region is balanced at certain levels of the recursion. This technique was initially used in the context of parallel and dynamic shortest-path algorithms \cite{Subramanian1995}, and the invariants it guarantees is often a necessity for algorithms for planar graphs \cite{???}. We state it here as a lemma:
\else
\paragraph{Decompositions of unweighted planar graphs.}
An $r$-division is a widely used decomposition of planar graphs into regions with small boundary. We use the $r$-divisions with a few holes as studied in~\cite{Klein2013}, which works for triangulated biconnected graphs:
\fi

\begin{Lem}{($r$-division with few holes for triangulated graphs~\cite{Klein2013})}
Let $G$ be a biconnected, triangulated $n$-vertex planar embedded graph, and let $0 < r \leq n$. $G$ can be decomposed into $\Theta(n/r)$ connected regions, each of which with $O(r)$ vertices and $O(\sqrt r)$ boundary vertices. Each region has a constant number of holes. Every boundary vertex lies on some hole, and each hole has $O(\sqrt r)$ vertices.\label{lem:r-division}
\end{Lem}

The fact that the boundaries of regions are small (only $O(\sqrt r)$ boundary vertices for a region with $r$ vertices) is the basis for many efficient algorithms and data structures for planar graphs. Unweighted planar graphs posses additional structure (in comparison to weighted planar graphs), which may also be useful algorithmically. See for example the unit-Monge property in~\cite{Abboud2018}, or the limited number of patterns~\cite{Wulff-Nilsen2013,Li2019}, which we use in this work. However, exploiting such additional structure in conjunction with a decomposition into regions with small boundaries has been elusive because of the seemingly technical requirement in Lemma~\ref{lem:r-division} that the graph be triangulated and biconnected.

Any graph can be triangulated and biconnected by adding to each face $f$ an artificial vertex and infinitely weighted artificial edges from the artificial vertex to each vertex of $V(f)$. This transformation preserves planarity and shortest paths, and ensures that the graph consists only of simple faces of size 3. However, the graph is no longer unweighted. We refer to an \textit{artificial} vertex (edge) of $G$ as a vertex (edge) which was added in the triangulation step, and a \textit{natural} vertex (edge) of $G$ as a vertex (edge) which is not artificial. In order to exploit the structure of the unweighted input graph we will remove the artificial edges and vertices after computing the decomposition using Lemma~\ref{lem:r-division}. 
On the one hand the graph is again unweighted. On the other hand, while the number of boundary vertices in each region remains $O(\sqrt r)$, the holes may now contain new non-boundary vertices, and the total size of the holes in each region may be $\Theta(r)$.
We note, however, that the deletion of artificial edges and vertices does not disconnect regions~\cite{Klein2013}. 
We therefore restate the decomposition lemma for unweighted graphs that are not necessarily triangulated or biconnected.
\begin{Lem}{($r$-division with few holes for non-triangulated graphs)}
Let $G$ be a $n$-vertex planar embedded graph $G$, and let $0 < r \leq n$. $G$ can be decomposed into $\Theta(n/r)$  connected regions, each of which with $O(r)$ vertices and $O(\sqrt r)$ boundary vertices. 
Each region has a constant number of holes, and each boundary vertex lies on some hole.
\label{lem:r-division-untriangulated}
\end{Lem}

\paragraph{Recursive $r$-divisions.}
Our second construction relies on a \textit{recursive $r$-division} which is a recursive decomposition of $G$ into $r$-divisions for varying values of $r$. Specifically, for a decreasing sequence $\mathbf{r} = r_1, r_2, \hdots $, where $n \geq r_1>r_2 > \hdots \geq 1$, we want $r_i$-divisions for all $i=1,2,\hdots$, such that each region in the $r_i$ division is the union of regions in the $r_{i+1}$-division on the next level. We associate with the recursive $r$-division a decomposition tree, $\mathcal{T}_\mathbf{r}$, which is a rooted tree whose nodes correspond to the regions of the recursive decomposition of $G$. We will refer to nodes and their corresponding regions interchangeably. The root node corresponds to all of $G$. A node $x$ of $\mathcal{T}_\mathbf{r}$ at depth $i$ corresponds to a region of the $r_i$-division, and its children are the regions of the $r_{i+1}$-division whose union is the region corresponding to $x$.  
We denote by $\mathcal{T}_{\mathbf{r}}^i$ all the nodes at level $i$.  
It was shown in~\cite{Klein2013} that recursive $r$-divisions can be computed efficiently:
\begin{Lem}{(Recursive $r$-division)}
Given a biconnected, triangulated $n$-vertex planar graph $G$ and an exponentially decreasing sequence $\mathbf{r}=n \geq r_1,r_2,\hdots  \geq 1$, a decomposition tree, $\mathcal{T}_\mathbf{r}$ can be computed in linear time s.t $\mathcal{T}_{\mathbf{r}}^i$ corresponds to an $r_i$-division of $G$ with few holes for each $i$.
\label{lem:r-division-recursive}
\end{Lem}

\section{Patterns}\label{sec:patterns}
\begin{figure}[!h]
\centering
\includegraphics[width=0.6\textwidth]{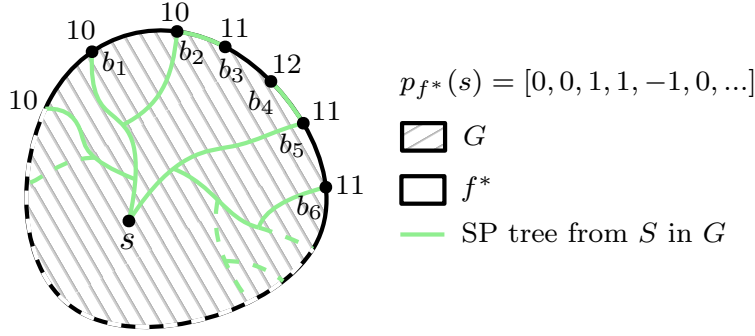}
\caption{Illustration of the pattern of the vertex $s$ w.r.t. $f^*$ in an undirected graph. In this case $f^*$ is the external face of the embedding. The numeric labels indicate the shortest path distances from $s$ to each $b_i$ where $b_i \in V(f^*)$ for $1 \leq i \leq 6$.}\label{fig:pattern}
\end{figure}

Both \cite{Wulff-Nilsen2013} and \cite{Li2019} introduce a notion of a ``distance tuple'' which can be thought of as a vector of shortest-path distances from a vertex to consecutive vertices of some hole. We introduce the following similar notion of a \textit{pattern} (See Figure \ref{fig:pattern} for an illustration): 

\begin{Def}{(Pattern)}
Let $G$ be a graph. Let $H$ be a subgraph of $G$. Let $u$ be a vertex in $H$, and let $\beta = b_0, b_1, \hdots, b_k$ be a path in $H$.
The \textit{pattern} of $u$ (w.r.t. $\beta$ in $H$) is a vector $p_{\beta,H}(u) \in \{-1,0,1\}^k$ satisfying 
$p_{\beta,H}(u)[i] = \dist{H}{u,b_i} - \dist{H}{u,b_{i-1}}$ for $1 \leq i \leq k$.
For a region $R$ in $G$, a hole $h$ of $R$, and a vertex $u \in \VV(h)$, we write $p_{h,G}(u)$ instead of $p_{w(h),G}(u)$.
\label{def:pattern}
\end{Def}

\begin{Def}{(pattern to vertex distance)} \label{def:p2v}
Let $R$ be a region in a graph $G$. Let $h$ be a hole of $R$. Let $b_0, b_1, \dots, b_k$ be the vertices of $w(h)$ in their cyclic order. 
Let $p$ be some pattern w.r.t. $h$ (i.e., $p=p_h(u)$ for some $u \in \VV(h)$). For a vertex $v \in R$ we define $\dist{G}{p,v}$ the \textit{distance between $p$ and $v$} to be $\min_{i=0}^k \left\{ \dist{G}{b_i,v} + \sum_{j=0}^i p[j]\right\}$.
\end{Def}

\begin{Lem} \label{lem:dist-pattern}
	Let $R$ be a region of a graph $G$. Let $h$ be a hole of $R$. For every $u \in \VV(h)$ and every $v\in R$,  $\dist{G}{u,v} = \dist{G}{u,b_0} + \dist{G}{p_h(u),v)}$.
\begin{proof}
By definition of pattern and by a telescoping sum, for every $0\leq i\leq k$, $\dist{G}{u,b_i} = \dist{G}{u,b_0} + \sum_{j=0}^i p[j]$. Let $b_\ell$ be any vertex of $w(h)$ on a shortest $u$-to-$v$ path ($b_\ell$ exists since $u \in  \VV(h)$ and $v \in R$). By choice of $b_\ell$, $\dist{G}{u,v} = \dist{G}{u,b_\ell} + \dist{G}{b_\ell,v} = \min_{0\leq i\leq k} \left\{\dist{G}{u,b_i} + \dist{G}{b_i,v}\right\} = \min _{0\leq i\leq k} \left\{\dist{G}{u,b_0} + \sum_{j=0}^i p[j] + \dist{G}{b_i,v}\right\} = \dist{G}{u,b_0} + \dist{G}{p,v}$.
\end{proof}
\end{Lem}

\paragraph{Bounding the number of patterns} As mentioned, 
In a recent paper, Li and Parter~\cite{Li2019} achieve improved bounds for diameter computation for planar graphs by showing that in unweighted undirected planar graphs the number of patterns is quite small. More specifically, they show that the VC-dimension of a set corresponding to all patterns is at most 3. By the Sauer-Shelah lemma~\cite{Sauer1972}, this implies that the number of distinct patterns w.r.t. a face $f$ is in $O(|S|^3)$.
Their result is stated in the following lemma:

\begin{Lem}{(Pattern compression)~\cite{Li2019}}
Let $G'=(V,E)$ be an $n$-vertex unweighted undirected planar graph, let $f$ be a 
face in $G'$, and let $S$ be a set of consecutive vertices on $f$. Then the number of distinct patterns w.r.t. $S$, $|\cup_{u \in V} \left\{ p_{S,G'}(u) \right\}|$, is bounded by $O(|S|^3)$. \label{lem:LPcompression}
\end{Lem}

We observe that the bound of Lemma~\ref{lem:LPcompression} also holds for patterns w.r.t. the entire set of vertices on a hole $h$ of a region $R$ even when distances are defined in the entire graph $G$. 
\begin{Cor}
Let $R$ be a region in an $n$-vertex unweighted undirected planar graph $G$, and let $h$ be a hole of $R$. 
Then  the number of distinct patterns w.r.t. $h$, $|\cup_{u \in \VV[h]} \left\{ p_{h,G}(u) \right\}|$, is bounded by $O(|h|^3)$. \label{lem:cycle-compression}	

\begin{proof}
	Since $h$ is a hole of $R$, $h$ is a face of $G-(R-h)$.
By Lemma~\ref{lem:LPcompression},  $|\cup_{v \in \VV[h]} \left\{ p_{h,G-(R-h)}(v) \right\}| = O(|h|^3)$. The corollary follows since for every two vertices $u,u' \in \VV(h)$, $p_{h,G-(R-h)}(u) = p_{h,G-(R-h)}(u')$ implies $p_{h,G}(u) = p_{h,G}(u')$.
\end{proof}
\end{Cor}

For the remainder of the paper we only deal with distances in $G$ and with patterns in $G$, so we will omit the subscript $G$, and write $\dist{}{\cdot,\cdot}$ and $p_h(\cdot)$ instead of $\dist{}{\cdot,\cdot}$ and $p_{h,G}(\cdot)$. 

\section{$O(n^{7/4})$ space distance oracle} \label{sec:construction-1}
Before presenting our main result, we describe a simpler construction which yields a distance oracle with a larger space requirement of $O(n^{7/4})$ and $O(1)$ query time:

\paragraph{Preprocessing.} The preprocessing consists of computing an $r$-division $\mathcal R$ of $G$ with a parameter $r$ to be determined later. For every vertex $v$ of $G$ and every region $R$ of $\mathcal R$, we store the hole $h$ of $R$ s.t. $v$ is in $\VV(h)$. This requires $O(n \cdot n/r) = O(n^2/r)$ space.   

For every region $R \in \mathcal R$,  for every hole $h$ of $R$, we maintain the $O(r^3)$ patterns of the vertices in $\VV(h)$ w.r.t. $h$ as follows. 
Let $k$ denote the size of the boundary walk $w(h)$ of $h$. let $v_h$ be the canonical (i.e., first) vertex of $w(h)$. 
We maintain the patterns seen so far in a ternary tree $\patterntree$ whose edges are labeled by $\{-1,0,1\}$. The depth of $\patterntree$ is $k-1$, and the labels along each root-to-leaf path correspond to a unique pattern, which we associate with that leaf.  
For every vertex $v \in \VV(h)$, we compute 
the pattern $p_{h}(v)$ and we make sure that $p_{h}(v)$ is represented in the tree $\patterntree$ by adding the corresponding labeled edges that are not yet present in $\patterntree$.
After all the vertices in $\VV(h)$ were handled, the tree $\patterntree$ has $O(r^3)$ leaves.
For each leaf of $\patterntree$ with an associated pattern $p$, we compute and store (i) the distance from $p$ to each vertex of $R$. This requires $O(r^4)$ time and space for all leaves of $\patterntree$, so a total of $O(n/r \cdot r^4) = O(nr^3)$ space for storing all this information over all regions.

For each vertex $v \in \VV(h)$ we store (ii) a pointer to (the leaf of $\patterntree$ that is associated with) the pattern $p_{h,G}(v)$, as well as (iii) the distance $\dist{}{v,v_h}$ between $v$ and the canonical vertex of $h$.
The total space required to store all these pointers and distances is $O(n \cdot n/r) = O(n^2/r)$.
  
To complete the preprocessing we also store (iv) for each region $R \in \mathcal R$, the distance $\dist{}{u,v}$ for all pairs of vertices $u,v \in R$. This takes $O(n/r \cdot r^2)$ additional space, which is dominated by the above terms.

The total space required by the oracle is thus $O(n^2/r) + O(nr^3)$. This is minimized for $r=n^{1/4}$, resulting in an $O(n^{7/4})$-space data structure. 

We note that once this information has been computed we no longer need to store the entire tree $\patterntree$. Rather, it suffices to only store just the list of leaves of $\patterntree$ and the distances stored with each of them. In particular, we no longer need to remember what is the actual pattern associated with each leaf, we only need to know the distances from this pattern to the vertices of the region $R$. In the current scheme this has no asymptotic effect on the size of the data structure, since each pattern is of size $O(r)$, and we anyway store the $O(r)$ distances from each pattern to all vertices of $R$. However, in the recursive scheme in the next section this observation will become useful.

\begin{figure}[!h]\label{fig:simple-query}
\centering
\includegraphics[width=0.6\textwidth]{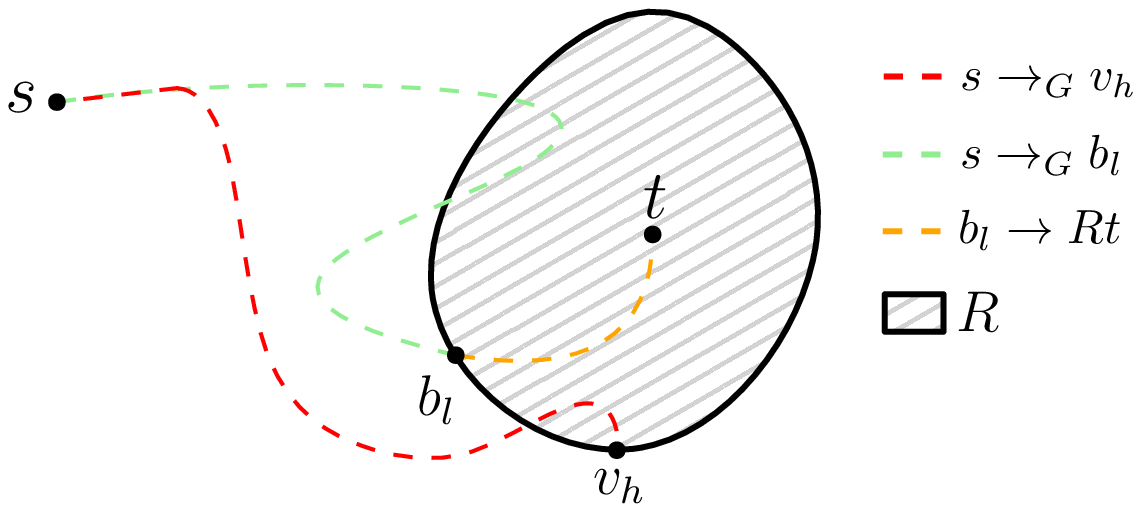}
\caption{Illustration of the query in Section \ref{sec:construction-1}. By Lemma~\ref{lem:dist-pattern} the query returns $\dist{}{s,b_0} + \dist{R}{p,t} = \dist{}{s,b_\ell} + \dist{R}{b_\ell,t} = \dist{}{s,t}$ where $b_\ell$ is some boundary vertex of $R$ on $s \leadsto t$.}
\end{figure}

\paragraph{Query.} To answer a query for the distance between vertices $s$ and $t$ we proceed as follows. If $s$ and $t$ are in the same regions, we simply return the distance $\dist{}{s,t}$ stored in item (iv). Otherwise, let $R$ be the region containing $t$, and let $h$ be the hole of $R$ such that $s \in \VV(h)$. Let $v_h$ be the canonical vertex of $h$. We return $\dist{}{s,v_h} + \dist{}{p_{h,G}(s),t}$. The correctness is immediate from Lemma~\ref{lem:dist-pattern}. We note that $\dist{}{s,v_h}$ is stored in item (iii), a pointer to $p_{h,G}(s)$ is stored in item (ii), and $\dist{}{p_{h,G}(s),t}$ is stored in item (i). The query is illustrated in Figure \ref{fig:simple-query}.

As aforementioned this oracle can be distributed into a distance labeling of size $O(n^{3/4})$ per label such that the distance between any two vertices $s,t$ can be computed in $O(1)$ time given just the labels of $s$ and $t$.

\section{$O(n^{5/3+\varepsilon})$ space distance oracle}\label{sec:construction-3}
A bottleneck in the above approach comes from having to store, for each pattern $p$ of a hole $h$ of a region $R$,  
the distances from $p$ to all vertices of $R$. Instead, we use a recursive $r$-division, in which we store for $p$, only the distances to the canonical vertex of a hole $h'$ of each child region $R'$ of $R$ instead of all the vertices in the region. For this information to be useful we also store the pattern induced by $p$ on the hole $h'$, which is defined as follows.

\begin{Def}{(Pattern induced by a pattern)}
Let $R$ be a region in a graph $G$. Let $h$ be a hole of $R$ and $p_h$ be a pattern of $h$ (w.r.t. a vertex or another pattern). Let $R'$ be a child region of $R$. 
Let $v_{h'} = b_0, b_1, \hdots, b_k$ be the vertices of the boundary walk of a hole of $h'$ of $R'$. 
The pattern induced by $p_h$ on $h'$ is the vector $p_{h'}$ satisfying $p_{h'}[i] = \dist{}{p,b_i} - \dist{}{p,b_{i-1}}$ for $1 \leq i \leq k$.
\label{def:pattern-pattern}
\end{Def}

\begin{Lem} \label{lem:p2p}
Consider the settings of Definition~\ref{def:pattern-pattern}. If $p_h = p_h(u)$ for some $u \in \VV(h)$, then $p_{h'} = p_{h'}(u)$.
\begin{proof}
By Lemma~\ref{lem:dist-pattern}, for every $0 \leq i \leq k$, $\dist{}{u,b_i} -\dist{}{u,v_h} = \dist{}{p,b_i}$. Hence for all $1 \leq i\leq k$, $p_{h'}[i] = \dist{}{p,b_i} - \dist{}{p,b_{i-1}}	 = \dist{}{u,b_i} -\dist{}{u,v_h} - (\dist{}{u,b_{i-1}} -\dist{}{u,v_h}) = \dist{}{u,b_i} - \dist{}{u,b_{i-1}}$, which is, by definition, $p_{h'}(u)[i]$.
\end{proof}

\end{Lem}

\paragraph{Preprocessing.}
We first compute a $\mathbf{r}=(r_0,r_1,\hdots,r_{k},r_{k+1})$-division of $G$ for $\mathbf{r}$ to be determined later, and denote by $\mathcal{T}_\mathbf{r}$ the associated decomposition tree. For convenience, we let $r_0=n$, $r_{k+1}=1$ and define $C(R)=\left\{ R' \; | \; R' \text{ is a child of } R \text{ in } \mathcal T_\mathbf{r} \right\}$. In the following we let $P_h$ denote the set $\{p_h(u) : u \in G\}$.
We store the following:

\begin{enumerate}
\item For each $u \in V(G)$ we 
store a list of regions $R_0 \supset R_1 \supset \dots \supset R_k$ containing $u$, where $R_i \in \mathcal T_\mathbf{r}^i$. (Recall that $\mathcal{T}_{\mathbf{r}}^i$ is the set of all nodes of $T_\mathbf{r}$ at level $i$).  
\item For each $u \in G$, for each $0 \leq i \leq k-1$, for each region $R \in \mathcal T_\mathbf{r}^i$ containing $u$, for each child region $R' \subset R$ at level-$(i+1)$, let $h$ be the hole of $R'$ such that $u \in \VV(h)$. 
We associate with the pair $(u,R')$ (i) a pointer to $p_h(u)$, (ii) the canonical vertex $v_h$, and (ii) the distance $\dist{}{u,v_h})$.   
\item For each $1 \leq i \leq k$, for each $R \in \mathcal T_\mathbf r^{i}$, for each hole $h$ in $R$, for each $p \in P_h$ and for each $R' \in C(R)$, let $h'$ be the hole of $R'$ such that $v_h \in \VV(h')$. 
We associate with the pair $(p,R')$ (i) a pointer to the pattern $p_{h'}(p)$ induced by $p$ on $h'$, (ii) the canonical vertex $v_{h'}$, and (iii) the distance $ \dist{}{p,v_{h'}})$. 
\end{enumerate}

\paragraph{Space analysis.} Storing 1 requires space $O(kn)$.
To bound the space for item 2, we note that the number of regions at level $i$ to which a vertes $u$ belongs is bounded by the degree of $u$. Since the average vertex degree in a planar graph is at most 6, the average number of regions at level $i$ to which $u$ belongs is at most 6. Each such region has $r_i/r_{i+1}$ subregions at level-$(i+1)$, so 
storing 2 requires space $O(n\sum_{i=0}^{k-1} r_i/r_{i+1})  = O(n^2/r) + O(n\sum_{i=1}^{k-1} r_i/r_{i+1})$.    
Storing 3 requires space $O(\sum_{i=1}^k (n/r_i) \cdot r_i^3 \cdot r_i/r_{i+1})=O(n \sum_{i=1}^k r_i^3/r_{i+1})$.
The total space is thus, $O(nk + n^2/r + n \sum_{i=1}^k r_i^3/r_{i+1})$.

\begin{algorithm}[b]
\caption{Query procedure for the $O(n^{5/3+\varepsilon})$ construction.}
\begin{algorithmic}[1]
\Procedure{Query}{$s,t$}
  \State{$i \gets$ the largest $i$ s.t. the region $R_i$ stored in item 1 for $t$ contains both $s$ and $t$}
  \State{$R_t \gets$ level $(i+1)$ region stored in item 1 for $t$}  
  \State{$(p,d) \gets$ the tuple associated with $(s,R_t)$}
  \State $i \leftarrow i+1$
  \While{$i\leq k$}
    \State{$R'_t \gets$ level $(i+1)$ subregion of $R_t$ stored in item 1 for $t$} 	
    \State{$(p',d') \gets$ the tuple associated with $(p,R'_t)$}
    \State $d \leftarrow d+d'$ ; $p \leftarrow p'$ ; $R_t \leftarrow R'_t$ ; $i \leftarrow i+1$
  \EndWhile
  \State{\Return{$d$}}
\EndProcedure
\end{algorithmic}
\label{alg:final-query}
\end{algorithm}

\paragraph{Query.} Algorithm \ref{alg:final-query} show pseudocode describing the query procedure. To process a query $\dist{}{s,t}$ the query procedure first determines the largest value $i$ for which $s$ and $t$ belong to the same region in $\mathcal{T}_\mathbf{r}^i$. 
Note that such a region must always exists as the root of $\mathcal{T}_\mathbf{r}$ is all of $G$. 
This level can be found in $O(k)$ time by traversing $\mathcal{T}_\mathbf{r}$, starting from a leaf region containing $s$ and a leaf region containing $t$.

Let $R_t$ be the level-$(i+1)$ region stored for $t$ in item 1. Note that, $t\in R_t$, and, by choice of $i$, $s \notin R_t$. 
Hence, $s$ is in some hole $h$ of $R_t$.  
We retrieve the pattern $p_h(s)$ and the distance $\dist{}{s,v_h}$ associated with $(s,R_t)$ in item 2.
We then proceed iteratively "zooming" into increasingly smaller regions containing $t$. 

We show that the algorithm maintains the invariant that, at the beginning of each iteration, we have a level-$i$ region $R_t$  containing $t$, the variable $d$ stores $\dist{}{s,v_h}$, where $h$ is the hole of $R_t$ such that $s\in \VV(h)$, and the variable $p$ stores (a pointer) to the pattern $p_h(t)$. Thus, when we reach the singleton region containing $t$, the variable $d$ stores $\dist{}{s,t}$.

We have already established that the invariant is maintained just before the loop is entered for the first time.
In each iteration of the loop
we retrieve $R'_t$, a level-$(i+1)$ subregion or $R_t$ containing $t$ (available in item 1), and retrieve  $d' \leftarrow \dist{}{p,v_{h'}}$ and $p' \leftarrow p_{h'}(p)$ (associated with the pair $(p,R'_t)$ in item 3).
By Lemma~\ref{lem:dist-pattern}, $d + d' = \dist{}{s,v_h} + \dist{}{p_h(u),v_{h'}} = \dist{}{s,v_{h'}}$. By Lemma~\ref{lem:p2p}, $p' = p_{h'}(t)$. 
Hence, after the assignements in Line~9, the invariant is restored.

The time complexity of the query is clearly $O(k)$. 

\paragraph{Choosing parameters:} 
Recall that the space requirement is $O(nk + n^2/r + n \sum_{i=1}^k r_i^3/r_{i+1})$. Picking each $r_i$ s.t. $r_i/r_{i+1}=r_1^{\varepsilon}$ results in $r_k=\Theta(1)$ when $k=\Theta(1/\varepsilon)$, and in a query time of $O(1/\varepsilon)$. Choosing $r_1 = n^{1/3+\varepsilon}$, the total space used becomes
\begin{align*}
O\left( n\sum_{i=1}^k r_i^3/r_{i+1} \right)=O(nr_1^2 r_1^{\varepsilon})=O(n^{1+2/3+2\varepsilon+\varepsilon/3+\varepsilon^2})=O(n^{5/3+\varepsilon'})
\end{align*}
for a suitable choice of $\varepsilon'$.

One can decrease the sizes of regions more aggressively to get the query time of  $k=O(\log(1/\varepsilon))$ of Theorem 1.
To this end we choose $\mathbf{r}$ such that $r_i^3/r_{i+1} = n^{2/3+\varepsilon}$, and $r_1 = n^{1/3}$. 
Then the space requirement is $O(n^{5/3} + nkn^{2/3+\varepsilon}) =O(kn^{5/3+\varepsilon})$.
It is not hard to verify that one gets $r_i = O(n^{1/3- \varepsilon \frac{3^{i-2}-1}{2}})$, so  $r_k = O(1)$ with $k=O(\log(1/\varepsilon))$. 

As a last remark we note that the smallest interesting choice of $\varepsilon$ in Theorem~\ref{thm:main} is $\Theta(1/\log n)$, giving $O(n^{5/3})$ space and $O(\log\log n)$ query-time, which is a faster query-time than was previously known for this amount of space \cite{Cohen-Addad2017,Charalampopoulos2019a}.

\bibliographystyle{abbrv}

\end{document}